\documentclass[twocolumn,aip,jcp,reprint]{revtex4-1}
\usepackage{graphicx}
\usepackage[caption=false]{subfig}
\begin{document}
%%%

%%%%%General information

%%% Title 
\title{Tailoring superelasticity of soft magnetic materials}

%%% Authors
\author{Peet Cremer}
\email{pcremer@thphy.uni-duesseldorf.de}
\affiliation{Institut f{\"u}r Theoretische Physik II: Weiche Materie, 
Heinrich-Heine-Universit{\"a}t D{\"u}sseldorf, D-40225 D{\"u}sseldorf, Germany}
\author{Hartmut L{\"o}wen}
\affiliation{Institut f{\"u}r Theoretische Physik II: Weiche Materie, 
Heinrich-Heine-Universit{\"a}t D{\"u}sseldorf, D-40225 D{\"u}sseldorf, Germany}
\author{Andreas M. Menzel}
\email{menzel@thphy.uni-duesseldorf.de}
\affiliation{Institut f{\"u}r Theoretische Physik II: Weiche Materie, 
Heinrich-Heine-Universit{\"a}t D{\"u}sseldorf, D-40225 D{\"u}sseldorf, Germany}

%%% Date
\date{\today}

%%% pacs
%\pacs{82.70.Dd, 75.80.+q, 82.35.Np, 82.70.Gg} 

%%% Abstract
\begin{abstract}
Embedding magnetic colloidal particles in an elastic polymer matrix leads to smart soft materials that can reversibly be addressed from outside by external magnetic fields. 
We discover a pronounced nonlinear superelastic stress-strain behavior of such materials using numerical simulations. 
This behavior results from a combination of two stress-induced mechanisms: a detachment mechanism of embedded particle aggregates as well as a reorientation mechanism of magnetic moments. 
The superelastic regime can be reversibly tuned or even be switched on and off by external magnetic fields and thus be tailored during operation. 
Similarities to the superelastic behavior of shape-memory alloys suggest analogous applications, with the additional benefit of reversible switchability and a higher biocompatibility of soft materials.
\end{abstract}

%%% Maketitle
\maketitle

%%% Introduction
The term ``superelasticity'' expresses the capability of certain materials to perform huge elastic deformations that are completely reversible \cite{Obukhov1994_Macromolecules,Mullin2007_PhysRevLett}. 
It was initially introduced in the context of shape-memory alloys \cite{Otsuka2002_MRSBull,Otsuka2005_ProgMaterSci,Liu2013_SciRep}. 
These metallic materials can perform large recoverable deformations due to stress-induced phase transitions. 
A transition to a more elongated lattice structure accommodates an externally imposed extension. 
Typically, this transition shows up as a pronounced ``plateau-like'' regime on the corresponding stress-strain curve. 
On this plateau, the samples are heterogeneous with domains of already transitioned material. Then, only relatively small additional stress induces a huge additional deformation.
Smart material properties are observed  \cite{Wei1998_JMaterSci,Bellouard2008_MaterSciEngA,Sun2012_MaterDes,Jani2014_MaterDesign}: upon stress release, shape-memory alloys can reversibly find back to their initial state. 
They self-reliantly adapt their appearance to changed environmental conditions.

In the present letter, we demonstrate that an analogous phenomenological behavior can be realized for a very different class of materials, exploiting different underlying mechanisms. 
Moreover, we show that during operation the behavior can be reversibly tailored from outside by external magnetic fields. 
All of this is achieved by employing soft magnetic gels as working materials: colloidal magnetic particles embedded in a possibly swollen elastic polymer matrix \cite{Zrinyi1995_PolymGelsNetw}. 
Similarly to magnetic fluids \cite{McGrother1996_PhysRevLett,Osipov1997_JPhysAMathGen, Teixeira2000_JPhysCondensMatter,Huke2004_RepProgPhys, Klapp2005_JPhysCondensMatter,Holm2005_CurrOpinColloidInterfaceSci, Ilg2006_JPhysCondensMatter,Odenbach2013_MRSBull}, magnetic gels allow to reversibly adjust their material properties by external magnetic fields. 
In this way, switching the elastic properties \cite{Filipcsei2007_AdvPolymSci,Xu2011_SoftMatter,Ilg2013_SoftMatter,Han2013_IntJSolidsStruct,Stoll2014_JApplPolymSci} 
offers a route to construct readily tunable dampers \cite{Sun2008_PolymTest} or vibration absorbers \cite{Deng2006_SmartMaterStruct}, 
while the possibility to switch the shape \cite{Filipcsei2007_AdvPolymSci,Nguyen2010_MacromolChemPhys,Snyder2010_ActaMater,Gong2012_ApplPhysLett} 
allows application as soft actuators \cite{Zhou2005_SmartMaterStruct,Zimmermann2006_JPhysCondensMatter,Boese2012_JIntellMaterSystStruct}. 

Here, we show that magnetic gels due to the interplay between magnetic and elastic interactions likewise feature superelastic behavior: it is enabled by a detachment mechanism of embedded magnetic particle aggregates and by a reorientation mechanism of magnetic moments. 
Both mechanisms are stress-induced and respond to external magnetic fields. 
Therefore, superelasticity can be switched on and off, and also its magnitude and position on the stress-strain curve can be reversibly tailored during operation, as has been realized for some special metallic components \cite{Soederberg2005_SmartMaterStruct,Karaca2006_ActaMater}.
The superelastic behavior in our case covers a significantly larger strain regime. 
Furthermore, soft gel-like materials generally provide a larger deformability and higher degree of 
biocompatibility \cite{ElFeninat2002_AdvEngMater,Liu2007_JMaterChem,Sokolowski2007_BiomedMater, Leng2009_MRSBull,Behl2010_AdvMater} than metallic alloys.
This becomes particularly important for medical applications \cite{Li2013_AdvFunctMater}. 
There has been significant effort to transfer the properties of shape-memory alloys to soft materials \cite{Mohr2006_ProcNatlAcadSciUSA,Liu2007_JMaterChem,Leng2009_MRSBull,Behl2010_AdvMater}. 
Here we report on reversibly tailoring superelastic properties by external magnetic fields. 

We concentrate on anisotropic uniaxial magnetic gels \cite{Collin2003_MacromolRapidCommun,Bohlius2004_PhysRevE,Filipcsei2007_AdvPolymSci,Han2013_IntJSolidsStruct}. 
They are manufactured by applying a strong external magnetic field during preparation, which leads to the formation of oriented straight chain-like aggregates of embedded magnetic particles \cite{Zubarev2000_PhysRevE,Hynninen2005_PhysRevLett,Auernhammer2006_JChemPhys,Smallenburg2012_JPhysCondensMatter}. 
After subsequent chemical cross-linking of the embedding polymer network, the particle positions get permanently locked \cite{Frickel2011_JMaterChem}. 
We assume that the magnetic moments carried by the particles are free to reorient. 
First, for diameters up to 10--15 nm, this applies within the interior of each magnetic particle \cite{Neel1949_AnnGeopHys}. 
Second, this is possible when each particle as a whole is free to rotate \cite{Frickel2011_JMaterChem}, e.g., when the polymer is not completely cross-linked in the immediate particle vicinity \cite{Gundermann2014_SmartMaterStruct}.
Another example are yolk-shell particles with a magnetic core that can rotate within the shell \cite{Liu2012_JMaterChem,Okada2013_Langmuir}.
If reorientations of the magnetic moments are blocked, only the first of the two mechanisms described below is active.

%%% Simulation
We identify a superelastic stress-strain behavior of uniaxial magnetic gels by numerically investigating the following model system.
Identical spherical colloidal particles, each carrying a permanent magnetic dipole moment, are embedded in a continuous elastic matrix. 
The elastic deformation energy of the matrix is described by a standard nearly-incompressible Neo-Hookean model \cite{Hartmann2003_IntJSolidsStruct,supplemental}. 
We tessellate the matrix into sufficiently small tetrahedra by Delaunay triangulation \cite{Geuzaine2009_IntJNumerMethEng}. 
Each tetrahedron may deform affinely, increasing its elastic energy, from which we extract restoring forces on its delimiting nodes.
Nodes attached to surfaces of rigid embedded particles transmit forces and torques to these particles.
Energy minimization with respect to all node and particle positions, as well as all particle and dipole orientations, is performed (see supplemental material for technical details \cite{supplemental}).

\begin{figure}[htbp]%
	\subfloat{\label{fig1a}}% Dummy for labeling a
	\subfloat{\label{fig1b}}% Dummy for labeling b
	\subfloat{\label{fig1c}}% Dummy for labeling c
	\centering%
	\includegraphics[width = 1.0\columnwidth]{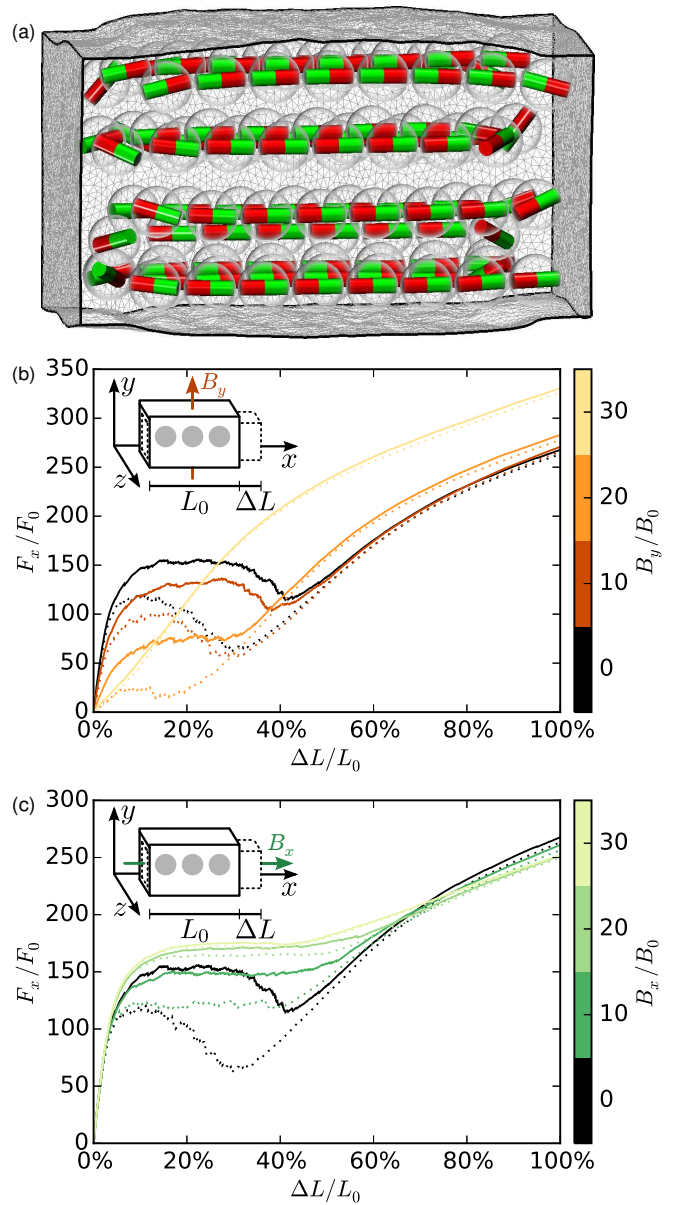}%
	\caption{%
		(Color online) 
		\protect\subref{fig1a} Snapshot of an equilibrated system of chain-like aggregates before stretching. 
		Small bar magnets in the embedded particles indicate dipole moments. 
		\protect\subref{fig1b} Uniaxial stress-strain behavior for stretching in chain direction, revealing a pronounced superelastic plateau-like nonlinearity, and adjustability by magnetic fields perpendicular to the chains. 
		\protect\subref{fig1c} Effects of magnetic fields parallel to the chains. 
		As in all subsequent figures, solid lines represent loading and dotted lines unloading, respectively, highlighting reversibility.
	}%
	\label{fig1}%
\end{figure}

Within our model, we study small three-dimensional systems, each containing $96$ magnetic particles (Fig.~\ref{fig1a}). 
For initialization, we arrange the particles in straight linear chain-like aggregates: each chain is one particle in diameter, but several equi-distanced particles in length that are separated by finite gaps filled with elastic material \cite{Han2013_IntJSolidsStruct,Annunziata2013_JChemPhys,Biller2014_JApplPhys, Gundermann2014_SmartMaterStruct,Menzel2014_JChemPhys}. 
The chains are initially aligned parallel to each other, but otherwise placed in a random non-overlapping way \cite{supplemental}. 
Through the presence of the rigid inclusions, the elastic modulus increases \cite{supplemental} by a factor of $\sim 7$.
Finally, the magnetic moments are switched on and the system is equilibrated, leading to an initial matrix deformation (Fig.~\ref{fig1a}).

We quasistatically stretch our systems along the chain direction. 
To impose a certain extension, the mesh nodes at two opposite system boundaries are displaced into opposite directions in small steps.
After equilibration during each step, the forces on the boundary nodes are measured. 
We check the reversibility of the induced total deformations by repeated loading and unloading cycles. 
Forces $F$ are measured in units of $F_0 = E R^2$, magnetic fields $B$ in units of $B_0 = \sqrt{\frac{\mu_0}{4\pi}E}$, and magnetic dipole moments $m$ in units of $m_0 = R^3\sqrt{\frac{4\pi}{\mu_0}E}$. 
Here, $E$ is the elastic modulus of the matrix, $R$ the particle radius, and $\mu_0$ the vacuum permeability. 
$L_0$ denotes the initial total length in stretching direction, $\Delta L$ the (absolute) elongation, and $\Delta L/L_0$ the elongational strain. 
We fix the material parameters by setting the Poisson ratio of the matrix to $\nu = 0.495$ and the magnetic moment to $m = 10 \, m_0$.

%%% Main results
Figs.~\ref{fig1b},\ref{fig1c} illustrate the resulting markedly nonlinear stress-strain behavior. 
First, the force to achieve a certain elongation steeply increases with the imposed strain. 
Then a pronounced superelastic nonlinearity follows. 
Since our measurements are strain-controlled and due to the finite size of our systems, we observe a regime of negative slope. 
A macroscopic sample in this region would become inhomogeneous leading to a plateau-like superelastic regime \cite{Wei1998_JMaterSci,Otsuka2005_ProgMaterSci} or likewise show negative slope under strain control \cite{Chernenko2003_JApplPhys}.   
In this area, a slight further increase in applied force induces a huge additional deformation. 
Remarkably, we can reversibly shift the non-linearity to smaller strains by a perpendicular external magnetic field (Fig.~\ref{fig1b}). 
High field strengths even switch off the non-linearity.
Furthermore, we can alter the shape of the plateau-like regime by a field in stretching direction (Fig.~\ref{fig1c}). 
At the end of the plateau, the stress-strain curve crosses over to a relatively constant intermediate slope. 

\begin{figure}[htbp]%
	\subfloat{\label{fig2a}}% Dummy for labeling a
	\subfloat{\label{fig2b}}% Dummy for labeling b
	\subfloat{\label{fig2c}}% Dummy for labeling c
	\subfloat{\label{fig2d}}% Dummy for labeling d
	\subfloat{\label{fig2e}}% Dummy for labeling e
	\subfloat{\label{fig2f}}% Dummy for labeling f
	\centering%
	\includegraphics[width = 0.96\columnwidth]{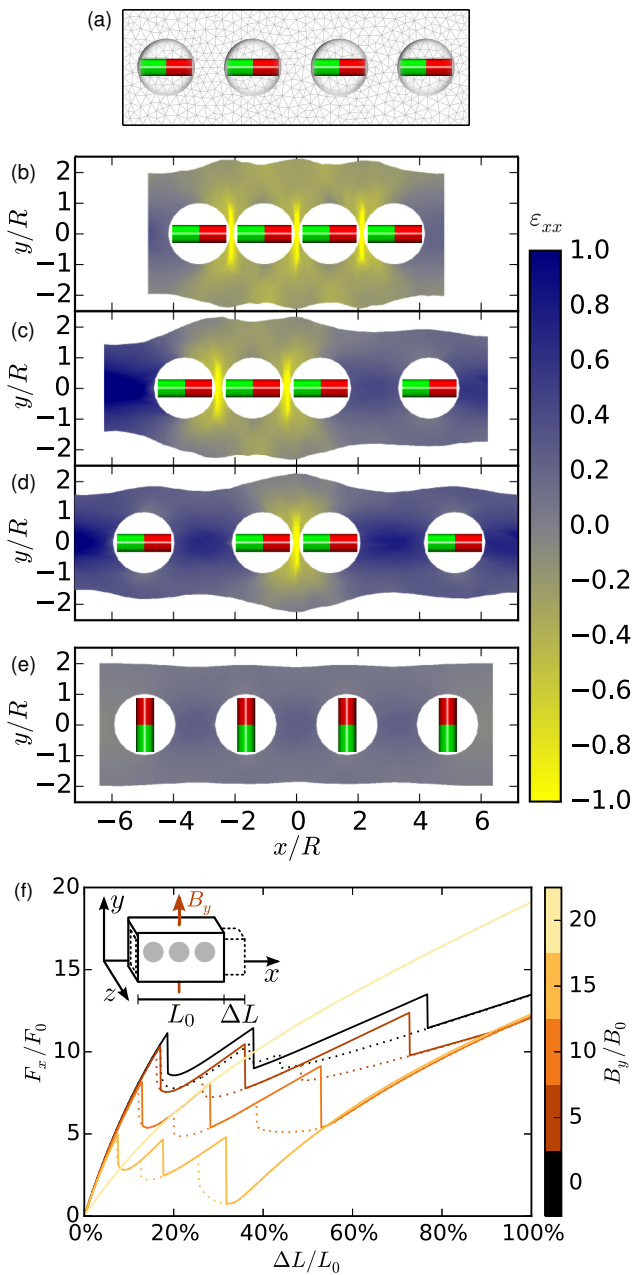}%
 	\caption{%
		(Color online) 
		Illustration of the detachment mechanism: 
		\protect\subref{fig2a} Cut through an initialized four-particle single-chain system, the elastic matrix not yet deformed. 
		\protect\subref{fig2b} Equilibrated state with pre-compressed gap material. 
		Color maps show the local matrix distortion along the longitudinal stretching direction, here illustrated using the so-called ``true strain'' $\varepsilon_{xx}$ \cite{Arghavani2011_IntJPlasticity}. 
		\protect\subref{fig2c} First and \protect\subref{fig2d} second detachment events of parts of the chain where magnetic energy barriers are overcome by longitudinally applied stretching forces. 
		\protect\subref{fig2e} Equilibrated unloaded state in the presence of a strong perpendicular magnetic field. 
		All magnetic moments are realigned, thus there is no pre-compression and no magnetic energy barrier. 
		\protect\subref{fig2f} Stress-strain curves for various perpendicular field strengths. 
		Each spike corresponds to a detachment event where a magnetic energy barrier is overcome. 
		An increasing perpendicular magnetic field lowers the detachment threshold, until superelasticity is switched off. 
	}%
	\label{fig2}%
\end{figure}

\begin{figure*}[htbp]%
	\subfloat{\label{fig3a}}% Dummy for labeling a
	\subfloat{\label{fig3b}}% Dummy for labeling b
	\subfloat{\label{fig3c}}% Dummy for labeling c
	\centering%
	\includegraphics[width = 0.92\textwidth]{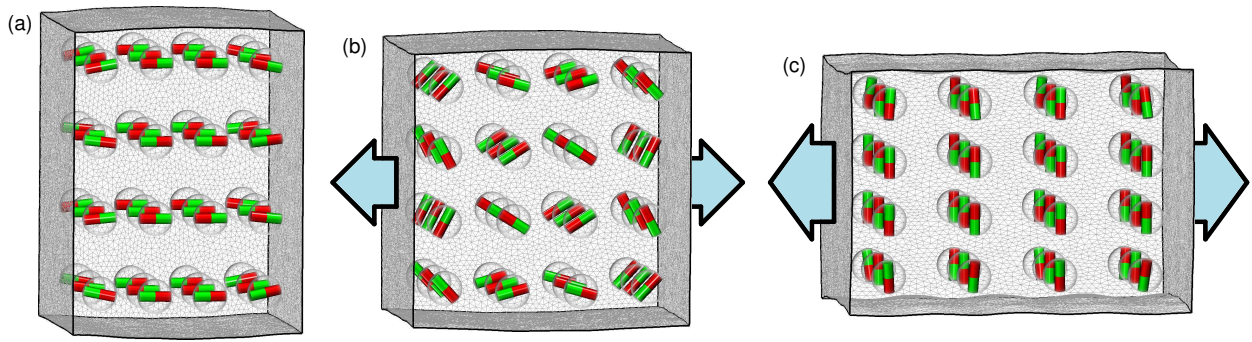}%
	\caption{%
		(Color online) 
		Illustration of the reorientation mechanism: snapshots of a stretched regular cuboid-lattice system (here $m = 8 \, m_0$). 
		\protect\subref{fig3a} Initially, the particles are closest in stretching direction (horizontal), their magnetic moments aligning along this direction. 
		Stretching the system increases distances along and (due to volume preservation) decreases distances perpendicular to the stretching direction. 
		\protect\subref{fig3b} When these distances become approximately equal, no particular orientation is preferred. 
		\protect\subref{fig3c} Stretching the system further makes a perpendicular direction the preferred one; the orientations of the magnetic moments ``flip''.
	}%
	\label{fig3}%
\end{figure*}

\begin{figure}[htbp]%
	\subfloat{\label{fig4a}}% Dummy for labeling a
	\subfloat{\label{fig4b}}% Dummy for labeling b
	\centering%
	\includegraphics[width = 1.0\columnwidth]{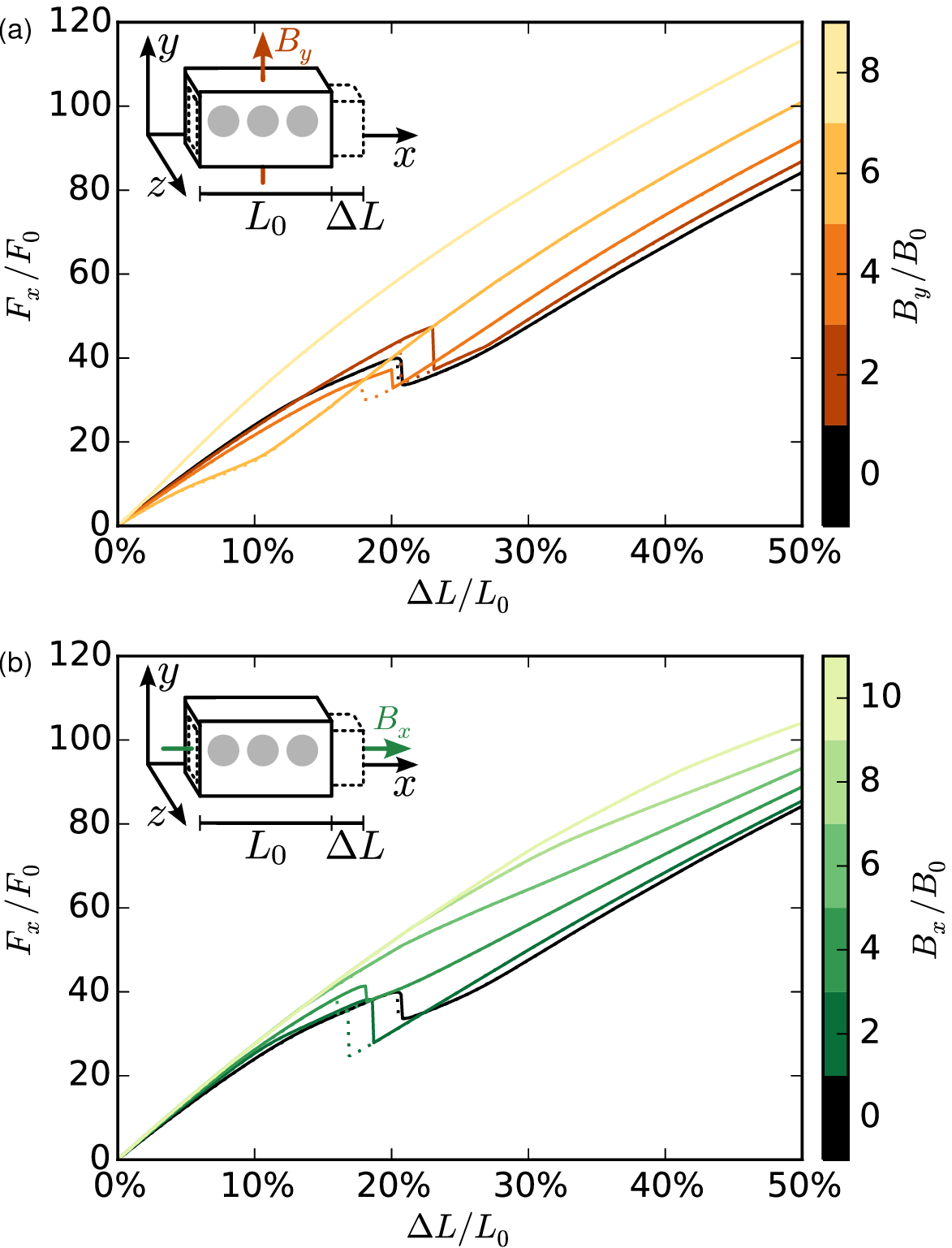}%
	\caption{%
		(Color online) 
		Stress-strain curves of the cuboid-lattice system showing a steep drop as reorientation (flipping) occurs. 
		\protect\subref{fig4a} Perpendicular external magnetic fields shift this feature to lower strains, while \protect\subref{fig4b} parallel external magnetic fields smear out the nonlinearity and postpone flipping to higher strains.
	}%
	\label{fig4}%
\end{figure}

%%% Chain detachment mechanism
We found that a combination of two effects allows for this adjustable superelastic behavior: 
a stress-induced detachment mechanism of the individual chain-like aggregates (Fig.~\ref{fig2}) plus a reorientation mechanism of the magnetic moments (Figs.~\ref{fig3},\ref{fig4}).
To illustrate the first one, Fig~\ref{fig2a} shows the initialized state of an example chain system. 
Switching on magnetic interactions, the particles attract each other and ``pre-compress'' the elastic gap material (Fig~\ref{fig2b}). 
Now the particles are located at a small distance from each other, its inverse cube setting the dipolar interaction scale. 
The initial steep increase of the stress-strain curve (Fig.~\ref{fig2f}) reflects these strong magnetic interactions. 
Once the magnetic barrier is overcome, a small further increase in stretching force is sufficient to detach part of the chain from the remainder (Fig.~\ref{fig2c}). 
Such events suddenly elongate the system and lead to spikes in the stress-strain curve (Fig.~\ref{fig2f}). 
They repeatedly occur (Fig.~\ref{fig2d}) until all particles have detached from each other. 
In total, a spiky plateau appears (Fig.~\ref{fig2f}). It smoothens when averaged over different chains (Fig.~\ref{fig1}). 
This is an intra-chain effect. Inter-chain interactions are minor for experimentally reported particle fractions \cite{Borbath2012_SmartMaterStruct,Han2013_IntJSolidsStruct}.
Thus, the finite size of our model systems will not affect the significance for the overall material behavior. 

Now, the tunability by a perpendicular magnetic field becomes clear (Figs.~\ref{fig2e},\ref{fig2f}). 
Strong perpendicular fields align all dipoles in the perpendicular direction. 
There is no pre-compression (Fig.~\ref{fig2e}), hence no magnetic energy barrier for pulling the particles apart, thus no corresponding stress-strain nonlinearity.
Smaller magnetic fields do not significantly alter the dipole orientations in the pre-compressed state due to the strong magnetic interactions at short distances. 
Yet they affect the threshold for detachment (Fig.~\ref{fig2f}): the moments can rotate away from the chain axes when the particle separation increases.

%%% Reorientation mechanism
The second effect contributing to the superelastic nonlinearity results from stretching-induced reorientations of the magnetic dipole moments. 
For illustration, we explain it on a regular cuboid lattice arrangement (Fig.~\ref{fig3}).
Initially, the edges of the cuboid unit cells are shorter along the stretching direction (Fig.~\ref{fig3a}). 
Thus the dipoles align parallel to it. 
During elongation (Fig.~\ref{fig3b}), these distances increase. 
Simultaneously, due to overall volume preservation, the system contracts from the sides. 
This decreases the separation perpendicular to the stretching direction. 
When the edge lengths of the distorted unit cells become equal in both directions, there is no single favored dipole orientation left (Fig.~\ref{fig3b}). 
Further stretching makes the dipoles rotate into the plane perpendicular to elongation (Fig.~\ref{fig3c}). 
Thus, during the overall process, the magnetic dipoles reorient (``flip''). 
This flipping effect is sensitive to the spatial particle arrangement. 
Yet, it likewise appeared in all of our investigated uniaxial systems. 
As becomes obvious from Fig.~\ref{fig3b}, we can identify it by a vanishing orientational order of the magnetic moments (see supplemental material for a quantitative evaluation \cite{supplemental}).  
A similar orientational analysis for the separation vectors between all nearest-neighbor particles demonstrates that the mechanism is indeed triggered by the distance changes mentioned above \cite{supplemental}. 

The flipping process is reflected by a steep step in the corresponding stress-strain curve (Fig.~\ref{fig4}; see supplemental material for a simplified energetic model including larger system sizes \cite{supplemental}).
Also this flipping contribution to the superelastic stress-strain behavior can be tuned from outside. 
Through a perpendicular magnetic field, flipping can be shifted to lower strains or be switched off completely, which largely eliminates the corresponding nonlinearity in the stress-strain curve (Fig.~\ref{fig4a}). 
However, also a parallel field has a significant influence: it can postpone flipping to larger deformations and smear out the connected stress-strain nonlinearity (Fig.~\ref{fig4b}). 
In Fig.~\ref{fig1c} it is the dip in the stress-strain curve that is mainly connected to the flipping mechanism and can be switched off by the parallel field. 
The steep jumps of Fig.~\ref{fig4} appear more rounded in Fig.~\ref{fig1b},\ref{fig1c} because all stress-strain curves in Fig.~\ref{fig1} were obtained by averaging over twenty characteristic numerical realizations.
For larger system sizes, the curves would appear still more rounded.

%%% Possible Experimental realization
To amplify the effects and to realize our assumed value of $m = 10 \, m_0$, strong magnetic moments and soft elastic matrices are preferred.
A possible route is to use particles made of a material of high remanent magnetization, for example NdFeB \cite{Kramarenko2015_SmartMaterStruct} (more than $2 \times 10^5 \, \textrm{A}/\textrm{m}$).
Soft elastic matrices of $E \lesssim 10^3 \, \textrm{Pa}$ can be made of silicone \cite{Hoang2009_SmartMaterStruct,Chertovich2010_MacromolMaterEng,Stoll2014_JApplPolymSci} or polydimethylsiloxane \cite{Huang2015_SoftMatter}.
The problem is qualitatively invariant under rescaling all lengths by a characteristic dimension such as the particle radius $R$.
Thus, the particle size is not a critical factor.
Our calculations were perfomed for permanent reorientable magnetic moments. 
Yet, the reorientation effect could likewise be observed using soft magnetic particles that are magnetized by an external magnetic field.
Then the reorientation process must be performed ``manually'' by switching the external magnetic field direction.

%%% Conclusion
In conclusion, we have identified a superelastic stress-strain behavior of soft uniaxial magnetic gels. 
These materials consist of chain-like aggregates of magnetic colloidal particles embedded in a soft elastic polymer matrix.
Stretching the systems in chain direction reveals a significant nonlinearity in the stress-strain curve.
In this regime, the systems can be strongly deformed with hardly any additional load necessary. 
Two underlying stress-induced mechanisms were identified: a detachment mechanism of the embedded chain-like aggregates and a reorientation mechanism of the magnetic moments. 
Both processes are reversible upon stress release, in analogy to the superelastic behavior of shape-memory alloys.   
As an additional benefit, the superelastic properties in the present case can be reversibly switched during operation by an external magnetic field. 

These nonlinear stress-strain properties open the pathway to numerous applications. 
The on-demand tunable deformability could be used for easily applicable packagings or gaskets that become rigid when an external magnetic field is switched off. 
Under pre-stress, external magnetic fields can trigger significant deformations, qualifying the materials for the use as soft actuators.
Combined with an increased biocompatibility, these concepts should be transferable to medical applications such as quick wound dressings, artificial muscles, or tunable implants.
Finally, combinations of magnetic gels with other materials can bestow the tunable superelastic properties on the resulting composite.

%%% Acknowledgments
\begin{acknowledgments}
The authors thank the Deutsche Forschungsgemeinschaft for support of this work through the priority program SPP 1681. 
\end{acknowledgments}

%%%Bibliography
%\bibliography{references}

\begin{thebibliography}{64}%
\makeatletter
\providecommand \@ifxundefined [1]{%
 \@ifx{#1\undefined}
}%
\providecommand \@ifnum [1]{%
 \ifnum #1\expandafter \@firstoftwo
 \else \expandafter \@secondoftwo
 \fi
}%
\providecommand \@ifx [1]{%
 \ifx #1\expandafter \@firstoftwo
 \else \expandafter \@secondoftwo
 \fi
}%
\providecommand \natexlab [1]{#1}%
\providecommand \enquote  [1]{``#1''}%
\providecommand \bibnamefont  [1]{#1}%
\providecommand \bibfnamefont [1]{#1}%
\providecommand \citenamefont [1]{#1}%
\providecommand \href@noop [0]{\@secondoftwo}%
\providecommand \href [0]{\begingroup \@sanitize@url \@href}%
\providecommand \@href[1]{\@@startlink{#1}\@@href}%
\providecommand \@@href[1]{\endgroup#1\@@endlink}%
\providecommand \@sanitize@url [0]{\catcode `\\12\catcode `\$12\catcode
  `\&12\catcode `\#12\catcode `\^12\catcode `\_12\catcode `\%12\relax}%
\providecommand \@@startlink[1]{}%
\providecommand \@@endlink[0]{}%
\providecommand \url  [0]{\begingroup\@sanitize@url \@url }%
\providecommand \@url [1]{\endgroup\@href {#1}{\urlprefix }}%
\providecommand \urlprefix  [0]{URL }%
\providecommand \Eprint [0]{\href }%
\providecommand \doibase [0]{http://dx.doi.org/}%
\providecommand \selectlanguage [0]{\@gobble}%
\providecommand \bibinfo  [0]{\@secondoftwo}%
\providecommand \bibfield  [0]{\@secondoftwo}%
\providecommand \translation [1]{[#1]}%
\providecommand \BibitemOpen [0]{}%
\providecommand \bibitemStop [0]{}%
\providecommand \bibitemNoStop [0]{.\EOS\space}%
\providecommand \EOS [0]{\spacefactor3000\relax}%
\providecommand \BibitemShut  [1]{\csname bibitem#1\endcsname}%
\let\auto@bib@innerbib\@empty
%</preamble>
\bibitem [{\citenamefont {Obukhov}, \citenamefont {Rubinstein},\ and\
  \citenamefont {Colby}(1994)}]{Obukhov1994_Macromolecules}%
  \BibitemOpen
  \bibfield  {author} {\bibinfo {author} {\bibfnamefont {S.~P.}\ \bibnamefont
  {Obukhov}}, \bibinfo {author} {\bibfnamefont {M.}~\bibnamefont {Rubinstein}},
  \ and\ \bibinfo {author} {\bibfnamefont {R.~H.}\ \bibnamefont {Colby}},\
  }\href@noop {} {\bibfield  {journal} {\bibinfo  {journal} {Macromolecules}\
  }\textbf {\bibinfo {volume} {27}},\ \bibinfo {pages} {3191} (\bibinfo {year}
  {1994})}\BibitemShut {NoStop}%
\bibitem [{\citenamefont {Mullin}\ \emph {et~al.}(2007)\citenamefont {Mullin},
  \citenamefont {Deschanel}, \citenamefont {Bertoldi},\ and\ \citenamefont
  {Boyce}}]{Mullin2007_PhysRevLett}%
  \BibitemOpen
  \bibfield  {author} {\bibinfo {author} {\bibfnamefont {T.}~\bibnamefont
  {Mullin}}, \bibinfo {author} {\bibfnamefont {S.}~\bibnamefont {Deschanel}},
  \bibinfo {author} {\bibfnamefont {K.}~\bibnamefont {Bertoldi}}, \ and\
  \bibinfo {author} {\bibfnamefont {M.~C.}\ \bibnamefont {Boyce}},\ }\href@noop
  {} {\bibfield  {journal} {\bibinfo  {journal} {Phys. Rev. Lett.}\ }\textbf
  {\bibinfo {volume} {99}},\ \bibinfo {pages} {084301} (\bibinfo {year}
  {2007})}\BibitemShut {NoStop}%
\bibitem [{\citenamefont {Otsuka}\ and\ \citenamefont
  {Kakeshita}(2002)}]{Otsuka2002_MRSBull}%
  \BibitemOpen
  \bibfield  {author} {\bibinfo {author} {\bibfnamefont {K.}~\bibnamefont
  {Otsuka}}\ and\ \bibinfo {author} {\bibfnamefont {T.}~\bibnamefont
  {Kakeshita}},\ }\href@noop {} {\bibfield  {journal} {\bibinfo  {journal} {MRS
  Bull.}\ }\textbf {\bibinfo {volume} {27}},\ \bibinfo {pages} {91} (\bibinfo
  {year} {2002})}\BibitemShut {NoStop}%
\bibitem [{\citenamefont {Otsuka}\ and\ \citenamefont
  {Ren}(2005)}]{Otsuka2005_ProgMaterSci}%
  \BibitemOpen
  \bibfield  {author} {\bibinfo {author} {\bibfnamefont {K.}~\bibnamefont
  {Otsuka}}\ and\ \bibinfo {author} {\bibfnamefont {X.}~\bibnamefont {Ren}},\
  }\href@noop {} {\bibfield  {journal} {\bibinfo  {journal} {Prog. Mater.
  Sci.}\ }\textbf {\bibinfo {volume} {50}},\ \bibinfo {pages} {511} (\bibinfo
  {year} {2005})}\BibitemShut {NoStop}%
\bibitem [{\citenamefont {Liu}\ \emph {et~al.}(2013)\citenamefont {Liu},
  \citenamefont {Wang}, \citenamefont {Hao}, \citenamefont {Wang},
  \citenamefont {Nie}, \citenamefont {Wang}, \citenamefont {Ren}, \citenamefont
  {Lu}, \citenamefont {Wang}, \citenamefont {Wang}, \citenamefont {Hui},
  \citenamefont {Lu}, \citenamefont {Kim},\ and\ \citenamefont
  {Yang}}]{Liu2013_SciRep}%
  \BibitemOpen
  \bibfield  {author} {\bibinfo {author} {\bibfnamefont {J.-P.}\ \bibnamefont
  {Liu}}, \bibinfo {author} {\bibfnamefont {Y.-D.}\ \bibnamefont {Wang}},
  \bibinfo {author} {\bibfnamefont {Y.-L.}\ \bibnamefont {Hao}}, \bibinfo
  {author} {\bibfnamefont {Y.}~\bibnamefont {Wang}}, \bibinfo {author}
  {\bibfnamefont {Z.-H.}\ \bibnamefont {Nie}}, \bibinfo {author} {\bibfnamefont
  {D.}~\bibnamefont {Wang}}, \bibinfo {author} {\bibfnamefont {Y.}~\bibnamefont
  {Ren}}, \bibinfo {author} {\bibfnamefont {Z.-P.}\ \bibnamefont {Lu}},
  \bibinfo {author} {\bibfnamefont {J.}~\bibnamefont {Wang}}, \bibinfo {author}
  {\bibfnamefont {H.}~\bibnamefont {Wang}}, \bibinfo {author} {\bibfnamefont
  {X.}~\bibnamefont {Hui}}, \bibinfo {author} {\bibfnamefont {N.}~\bibnamefont
  {Lu}}, \bibinfo {author} {\bibfnamefont {M.~J.}\ \bibnamefont {Kim}}, \ and\
  \bibinfo {author} {\bibfnamefont {R.}~\bibnamefont {Yang}},\ }\href@noop {}
  {\bibfield  {journal} {\bibinfo  {journal} {Sci. Rep.}\ }\textbf {\bibinfo
  {volume} {3}},\ \bibinfo {pages} {2156} (\bibinfo {year} {2013})}\BibitemShut
  {NoStop}%
\bibitem [{\citenamefont {Wei}, \citenamefont {Sandstr\"om},\ and\
  \citenamefont {Miyazaki}(1998)}]{Wei1998_JMaterSci}%
  \BibitemOpen
  \bibfield  {author} {\bibinfo {author} {\bibfnamefont {Z.}~\bibnamefont
  {Wei}}, \bibinfo {author} {\bibfnamefont {R.}~\bibnamefont {Sandstr\"om}}, \
  and\ \bibinfo {author} {\bibfnamefont {S.}~\bibnamefont {Miyazaki}},\
  }\href@noop {} {\bibfield  {journal} {\bibinfo  {journal} {J. Mater. Sci.}\
  }\textbf {\bibinfo {volume} {33}},\ \bibinfo {pages} {3743} (\bibinfo {year}
  {1998})}\BibitemShut {NoStop}%
\bibitem [{\citenamefont {Bellouard}(2008)}]{Bellouard2008_MaterSciEngA}%
  \BibitemOpen
  \bibfield  {author} {\bibinfo {author} {\bibfnamefont {Y.}~\bibnamefont
  {Bellouard}},\ }\href@noop {} {\bibfield  {journal} {\bibinfo  {journal}
  {Mater. Sci. Eng. A}\ }\textbf {\bibinfo {volume} {481-482}},\ \bibinfo
  {pages} {582} (\bibinfo {year} {2008})}\BibitemShut {NoStop}%
\bibitem [{\citenamefont {Sun}\ \emph {et~al.}(2012)\citenamefont {Sun},
  \citenamefont {Huang}, \citenamefont {Ding}, \citenamefont {Zhao},
  \citenamefont {Wang}, \citenamefont {Purnawali},\ and\ \citenamefont
  {Tang}}]{Sun2012_MaterDes}%
  \BibitemOpen
  \bibfield  {author} {\bibinfo {author} {\bibfnamefont {L.}~\bibnamefont
  {Sun}}, \bibinfo {author} {\bibfnamefont {W.}~\bibnamefont {Huang}}, \bibinfo
  {author} {\bibfnamefont {Z.}~\bibnamefont {Ding}}, \bibinfo {author}
  {\bibfnamefont {Y.}~\bibnamefont {Zhao}}, \bibinfo {author} {\bibfnamefont
  {C.}~\bibnamefont {Wang}}, \bibinfo {author} {\bibfnamefont {H.}~\bibnamefont
  {Purnawali}}, \ and\ \bibinfo {author} {\bibfnamefont {C.}~\bibnamefont
  {Tang}},\ }\href@noop {} {\bibfield  {journal} {\bibinfo  {journal} {Mater.
  Des.}\ }\textbf {\bibinfo {volume} {33}},\ \bibinfo {pages} {577} (\bibinfo
  {year} {2012})}\BibitemShut {NoStop}%
\bibitem [{\citenamefont {Jani}\ \emph {et~al.}(2014)\citenamefont {Jani},
  \citenamefont {Leary}, \citenamefont {Aleksandar},\ and\ \citenamefont
  {Gibson}}]{Jani2014_MaterDesign}%
  \BibitemOpen
  \bibfield  {author} {\bibinfo {author} {\bibfnamefont {J.~M.}\ \bibnamefont
  {Jani}}, \bibinfo {author} {\bibfnamefont {M.}~\bibnamefont {Leary}},
  \bibinfo {author} {\bibfnamefont {S.}~\bibnamefont {Aleksandar}}, \ and\
  \bibinfo {author} {\bibfnamefont {M.~A.}\ \bibnamefont {Gibson}},\
  }\href@noop {} {\bibfield  {journal} {\bibinfo  {journal} {Mater. Design}\
  }\textbf {\bibinfo {volume} {56}},\ \bibinfo {pages} {1078} (\bibinfo {year}
  {2014})}\BibitemShut {NoStop}%
\bibitem [{\citenamefont {Zr\'inyi}, \citenamefont {Barsi},\ and\ \citenamefont
  {B\"uki}(1995)}]{Zrinyi1995_PolymGelsNetw}%
  \BibitemOpen
  \bibfield  {author} {\bibinfo {author} {\bibfnamefont {M.}~\bibnamefont
  {Zr\'inyi}}, \bibinfo {author} {\bibfnamefont {L.}~\bibnamefont {Barsi}}, \
  and\ \bibinfo {author} {\bibfnamefont {A.}~\bibnamefont {B\"uki}},\
  }\href@noop {} {\bibfield  {journal} {\bibinfo  {journal} {Polym. Gels
  Netw.}\ }\textbf {\bibinfo {volume} {5}},\ \bibinfo {pages} {415} (\bibinfo
  {year} {1995})}\BibitemShut {NoStop}%
\bibitem [{\citenamefont {McGrother}\ and\ \citenamefont
  {Jackson}(1996)}]{McGrother1996_PhysRevLett}%
  \BibitemOpen
  \bibfield  {author} {\bibinfo {author} {\bibfnamefont {S.~C.}\ \bibnamefont
  {McGrother}}\ and\ \bibinfo {author} {\bibfnamefont {G.}~\bibnamefont
  {Jackson}},\ }\href@noop {} {\bibfield  {journal} {\bibinfo  {journal} {Phys.
  Rev. Lett.}\ }\textbf {\bibinfo {volume} {76}},\ \bibinfo {pages} {4183}
  (\bibinfo {year} {1996})}\BibitemShut {NoStop}%
\bibitem [{\citenamefont {Osipov}, \citenamefont {Teixeira},\ and\
  \citenamefont {Telo~da Gama}(1997)}]{Osipov1997_JPhysAMathGen}%
  \BibitemOpen
  \bibfield  {author} {\bibinfo {author} {\bibfnamefont {M.~A.}\ \bibnamefont
  {Osipov}}, \bibinfo {author} {\bibfnamefont {P.~I.~C.}\ \bibnamefont
  {Teixeira}}, \ and\ \bibinfo {author} {\bibfnamefont {M.~M.}\ \bibnamefont
  {Telo~da Gama}},\ }\href@noop {} {\bibfield  {journal} {\bibinfo  {journal}
  {J. Phys. A: Math. Gen.}\ }\textbf {\bibinfo {volume} {30}},\ \bibinfo
  {pages} {1953} (\bibinfo {year} {1997})}\BibitemShut {NoStop}%
\bibitem [{\citenamefont {Teixeira}, \citenamefont {Tavares},\ and\
  \citenamefont {Telo~da Gama}(2000)}]{Teixeira2000_JPhysCondensMatter}%
  \BibitemOpen
  \bibfield  {author} {\bibinfo {author} {\bibfnamefont {P.~I.~C.}\
  \bibnamefont {Teixeira}}, \bibinfo {author} {\bibfnamefont {J.}~\bibnamefont
  {Tavares}}, \ and\ \bibinfo {author} {\bibfnamefont {M.~M.}\ \bibnamefont
  {Telo~da Gama}},\ }\href@noop {} {\bibfield  {journal} {\bibinfo  {journal}
  {J. Phys.: Condens. Matter}\ }\textbf {\bibinfo {volume} {12}},\ \bibinfo
  {pages} {R411} (\bibinfo {year} {2000})}\BibitemShut {NoStop}%
\bibitem [{\citenamefont {Huke}\ and\ \citenamefont
  {L\"ucke}(2004)}]{Huke2004_RepProgPhys}%
  \BibitemOpen
  \bibfield  {author} {\bibinfo {author} {\bibfnamefont {B.}~\bibnamefont
  {Huke}}\ and\ \bibinfo {author} {\bibfnamefont {M.}~\bibnamefont {L\"ucke}},\
  }\href@noop {} {\bibfield  {journal} {\bibinfo  {journal} {Rep. Prog. Phys.}\
  }\textbf {\bibinfo {volume} {67}},\ \bibinfo {pages} {1731} (\bibinfo {year}
  {2004})}\BibitemShut {NoStop}%
\bibitem [{\citenamefont {Klapp}(2005)}]{Klapp2005_JPhysCondensMatter}%
  \BibitemOpen
  \bibfield  {author} {\bibinfo {author} {\bibfnamefont {S.~H.~L.}\
  \bibnamefont {Klapp}},\ }\href@noop {} {\bibfield  {journal} {\bibinfo
  {journal} {J. Phys.: Condens. Matter}\ }\textbf {\bibinfo {volume} {17}},\
  \bibinfo {pages} {R525} (\bibinfo {year} {2005})}\BibitemShut {NoStop}%
\bibitem [{\citenamefont {Holm}\ and\ \citenamefont
  {Weis}(2005)}]{Holm2005_CurrOpinColloidInterfaceSci}%
  \BibitemOpen
  \bibfield  {author} {\bibinfo {author} {\bibfnamefont {C.}~\bibnamefont
  {Holm}}\ and\ \bibinfo {author} {\bibfnamefont {J.-J.}\ \bibnamefont
  {Weis}},\ }\href@noop {} {\bibfield  {journal} {\bibinfo  {journal} {Curr.
  Opin. Colloid Interface Sci.}\ }\textbf {\bibinfo {volume} {10}},\ \bibinfo
  {pages} {133} (\bibinfo {year} {2005})}\BibitemShut {NoStop}%
\bibitem [{\citenamefont {Ilg}, \citenamefont {Coquelle},\ and\ \citenamefont
  {Hess}(2006)}]{Ilg2006_JPhysCondensMatter}%
  \BibitemOpen
  \bibfield  {author} {\bibinfo {author} {\bibfnamefont {P.}~\bibnamefont
  {Ilg}}, \bibinfo {author} {\bibfnamefont {E.}~\bibnamefont {Coquelle}}, \
  and\ \bibinfo {author} {\bibfnamefont {S.}~\bibnamefont {Hess}},\ }\href@noop
  {} {\bibfield  {journal} {\bibinfo  {journal} {J. Phys.: Condens. Matter}\
  }\textbf {\bibinfo {volume} {18}},\ \bibinfo {pages} {S2757} (\bibinfo {year}
  {2006})}\BibitemShut {NoStop}%
\bibitem [{\citenamefont {Odenbach}(2013)}]{Odenbach2013_MRSBull}%
  \BibitemOpen
  \bibfield  {author} {\bibinfo {author} {\bibfnamefont {S.}~\bibnamefont
  {Odenbach}},\ }\href@noop {} {\bibfield  {journal} {\bibinfo  {journal} {MRS
  Bull.}\ }\textbf {\bibinfo {volume} {38}},\ \bibinfo {pages} {921} (\bibinfo
  {year} {2013})}\BibitemShut {NoStop}%
\bibitem [{\citenamefont {Filipcsei}\ \emph {et~al.}(2007)\citenamefont
  {Filipcsei}, \citenamefont {Csetneki}, \citenamefont {Szil\'agyi},\ and\
  \citenamefont {Zr\'inyi}}]{Filipcsei2007_AdvPolymSci}%
  \BibitemOpen
  \bibfield  {author} {\bibinfo {author} {\bibfnamefont {G.}~\bibnamefont
  {Filipcsei}}, \bibinfo {author} {\bibfnamefont {I.}~\bibnamefont {Csetneki}},
  \bibinfo {author} {\bibfnamefont {A.}~\bibnamefont {Szil\'agyi}}, \ and\
  \bibinfo {author} {\bibfnamefont {M.}~\bibnamefont {Zr\'inyi}},\ }\href@noop
  {} {\bibfield  {journal} {\bibinfo  {journal} {Adv. Polym. Sci.}\ }\textbf
  {\bibinfo {volume} {206}},\ \bibinfo {pages} {137} (\bibinfo {year}
  {2007})}\BibitemShut {NoStop}%
\bibitem [{\citenamefont {Xu}\ \emph {et~al.}(2011)\citenamefont {Xu},
  \citenamefont {Gong}, \citenamefont {Xuan}, \citenamefont {Zhang},\ and\
  \citenamefont {Fan}}]{Xu2011_SoftMatter}%
  \BibitemOpen
  \bibfield  {author} {\bibinfo {author} {\bibfnamefont {Y.}~\bibnamefont
  {Xu}}, \bibinfo {author} {\bibfnamefont {X.}~\bibnamefont {Gong}}, \bibinfo
  {author} {\bibfnamefont {S.}~\bibnamefont {Xuan}}, \bibinfo {author}
  {\bibfnamefont {W.}~\bibnamefont {Zhang}}, \ and\ \bibinfo {author}
  {\bibfnamefont {Y.}~\bibnamefont {Fan}},\ }\href@noop {} {\bibfield
  {journal} {\bibinfo  {journal} {Soft Matter}\ }\textbf {\bibinfo {volume}
  {7}},\ \bibinfo {pages} {5246} (\bibinfo {year} {2011})}\BibitemShut
  {NoStop}%
\bibitem [{\citenamefont {Ilg}(2013)}]{Ilg2013_SoftMatter}%
  \BibitemOpen
  \bibfield  {author} {\bibinfo {author} {\bibfnamefont {P.}~\bibnamefont
  {Ilg}},\ }\href@noop {} {\bibfield  {journal} {\bibinfo  {journal} {Soft
  Matter}\ }\textbf {\bibinfo {volume} {9}},\ \bibinfo {pages} {3465} (\bibinfo
  {year} {2013})}\BibitemShut {NoStop}%
\bibitem [{\citenamefont {Han}, \citenamefont {Hong},\ and\ \citenamefont
  {Faidley}(2013)}]{Han2013_IntJSolidsStruct}%
  \BibitemOpen
  \bibfield  {author} {\bibinfo {author} {\bibfnamefont {Y.}~\bibnamefont
  {Han}}, \bibinfo {author} {\bibfnamefont {W.}~\bibnamefont {Hong}}, \ and\
  \bibinfo {author} {\bibfnamefont {L.~E.}\ \bibnamefont {Faidley}},\
  }\href@noop {} {\bibfield  {journal} {\bibinfo  {journal} {Int. J. Solids
  Struct.}\ }\textbf {\bibinfo {volume} {50}},\ \bibinfo {pages} {2281}
  (\bibinfo {year} {2013})}\BibitemShut {NoStop}%
\bibitem [{\citenamefont {Stoll}\ \emph {et~al.}(2014)\citenamefont {Stoll},
  \citenamefont {Mayer}, \citenamefont {Monkman},\ and\ \citenamefont
  {Shamonin}}]{Stoll2014_JApplPolymSci}%
  \BibitemOpen
  \bibfield  {author} {\bibinfo {author} {\bibfnamefont {A.}~\bibnamefont
  {Stoll}}, \bibinfo {author} {\bibfnamefont {M.}~\bibnamefont {Mayer}},
  \bibinfo {author} {\bibfnamefont {G.~J.}\ \bibnamefont {Monkman}}, \ and\
  \bibinfo {author} {\bibfnamefont {M.}~\bibnamefont {Shamonin}},\ }\href@noop
  {} {\bibfield  {journal} {\bibinfo  {journal} {J. Appl. Polym. Sci.}\
  }\textbf {\bibinfo {volume} {131}},\ \bibinfo {eid} {39793} (\bibinfo {year}
  {2014})}\BibitemShut {NoStop}%
\bibitem [{\citenamefont {Sun}\ \emph {et~al.}(2008)\citenamefont {Sun},
  \citenamefont {Gong}, \citenamefont {Jiang}, \citenamefont {Li},
  \citenamefont {Xu},\ and\ \citenamefont {Li}}]{Sun2008_PolymTest}%
  \BibitemOpen
  \bibfield  {author} {\bibinfo {author} {\bibfnamefont {T.}~\bibnamefont
  {Sun}}, \bibinfo {author} {\bibfnamefont {X.}~\bibnamefont {Gong}}, \bibinfo
  {author} {\bibfnamefont {W.}~\bibnamefont {Jiang}}, \bibinfo {author}
  {\bibfnamefont {J.}~\bibnamefont {Li}}, \bibinfo {author} {\bibfnamefont
  {Z.}~\bibnamefont {Xu}}, \ and\ \bibinfo {author} {\bibfnamefont
  {W.}~\bibnamefont {Li}},\ }\href@noop {} {\bibfield  {journal} {\bibinfo
  {journal} {Polym. Test.}\ }\textbf {\bibinfo {volume} {27}},\ \bibinfo
  {pages} {520} (\bibinfo {year} {2008})}\BibitemShut {NoStop}%
\bibitem [{\citenamefont {Deng}, \citenamefont {Gong},\ and\ \citenamefont
  {Wang}(2006)}]{Deng2006_SmartMaterStruct}%
  \BibitemOpen
  \bibfield  {author} {\bibinfo {author} {\bibfnamefont {H.-x.}\ \bibnamefont
  {Deng}}, \bibinfo {author} {\bibfnamefont {X.-l.}\ \bibnamefont {Gong}}, \
  and\ \bibinfo {author} {\bibfnamefont {L.-h.}\ \bibnamefont {Wang}},\
  }\href@noop {} {\bibfield  {journal} {\bibinfo  {journal} {Smart Mater.
  Struct.}\ }\textbf {\bibinfo {volume} {15}},\ \bibinfo {pages} {N111}
  (\bibinfo {year} {2006})}\BibitemShut {NoStop}%
\bibitem [{\citenamefont {Nguyen}\ and\ \citenamefont
  {Ramanujan}(2010)}]{Nguyen2010_MacromolChemPhys}%
  \BibitemOpen
  \bibfield  {author} {\bibinfo {author} {\bibfnamefont {V.~Q.}\ \bibnamefont
  {Nguyen}}\ and\ \bibinfo {author} {\bibfnamefont {R.~V.}\ \bibnamefont
  {Ramanujan}},\ }\href@noop {} {\bibfield  {journal} {\bibinfo  {journal}
  {Macromol. Chem. Phys.}\ }\textbf {\bibinfo {volume} {211}},\ \bibinfo
  {pages} {618} (\bibinfo {year} {2010})}\BibitemShut {NoStop}%
\bibitem [{\citenamefont {Snyder}, \citenamefont {Nguyen},\ and\ \citenamefont
  {Ramanujan}(2010)}]{Snyder2010_ActaMater}%
  \BibitemOpen
  \bibfield  {author} {\bibinfo {author} {\bibfnamefont {R.}~\bibnamefont
  {Snyder}}, \bibinfo {author} {\bibfnamefont {V.}~\bibnamefont {Nguyen}}, \
  and\ \bibinfo {author} {\bibfnamefont {R.}~\bibnamefont {Ramanujan}},\
  }\href@noop {} {\bibfield  {journal} {\bibinfo  {journal} {Acta Mater.}\
  }\textbf {\bibinfo {volume} {58}},\ \bibinfo {pages} {5620} (\bibinfo {year}
  {2010})}\BibitemShut {NoStop}%
\bibitem [{\citenamefont {Gong}, \citenamefont {Liao},\ and\ \citenamefont
  {Xuan}(2012)}]{Gong2012_ApplPhysLett}%
  \BibitemOpen
  \bibfield  {author} {\bibinfo {author} {\bibfnamefont {X.}~\bibnamefont
  {Gong}}, \bibinfo {author} {\bibfnamefont {G.}~\bibnamefont {Liao}}, \ and\
  \bibinfo {author} {\bibfnamefont {S.}~\bibnamefont {Xuan}},\ }\href@noop {}
  {\bibfield  {journal} {\bibinfo  {journal} {Appl. Phys. Lett.}\ }\textbf
  {\bibinfo {volume} {100}},\ \bibinfo {eid} {211909} (\bibinfo {year}
  {2012})}\BibitemShut {NoStop}%
\bibitem [{\citenamefont {Zhou}\ and\ \citenamefont
  {Wang}(2005)}]{Zhou2005_SmartMaterStruct}%
  \BibitemOpen
  \bibfield  {author} {\bibinfo {author} {\bibfnamefont {G.}~\bibnamefont
  {Zhou}}\ and\ \bibinfo {author} {\bibfnamefont {Q.}~\bibnamefont {Wang}},\
  }\href@noop {} {\bibfield  {journal} {\bibinfo  {journal} {Smart Mater.
  Struct.}\ }\textbf {\bibinfo {volume} {14}},\ \bibinfo {pages} {504}
  (\bibinfo {year} {2005})}\BibitemShut {NoStop}%
\bibitem [{\citenamefont {Zimmermann}\ \emph {et~al.}(2006)\citenamefont
  {Zimmermann}, \citenamefont {Naletova}, \citenamefont {Zeidis}, \citenamefont
  {B\"ohm},\ and\ \citenamefont {Kolev}}]{Zimmermann2006_JPhysCondensMatter}%
  \BibitemOpen
  \bibfield  {author} {\bibinfo {author} {\bibfnamefont {K.}~\bibnamefont
  {Zimmermann}}, \bibinfo {author} {\bibfnamefont {V.~A.}\ \bibnamefont
  {Naletova}}, \bibinfo {author} {\bibfnamefont {I.}~\bibnamefont {Zeidis}},
  \bibinfo {author} {\bibfnamefont {V.}~\bibnamefont {B\"ohm}}, \ and\ \bibinfo
  {author} {\bibfnamefont {E.}~\bibnamefont {Kolev}},\ }\href@noop {}
  {\bibfield  {journal} {\bibinfo  {journal} {J. Phys.: Condens. Matter}\
  }\textbf {\bibinfo {volume} {18}},\ \bibinfo {pages} {S2973} (\bibinfo {year}
  {2006})}\BibitemShut {NoStop}%
\bibitem [{\citenamefont {B\"ose}, \citenamefont {Rabindranath},\ and\
  \citenamefont {Ehrlich}(2012)}]{Boese2012_JIntellMaterSystStruct}%
  \BibitemOpen
  \bibfield  {author} {\bibinfo {author} {\bibfnamefont {H.}~\bibnamefont
  {B\"ose}}, \bibinfo {author} {\bibfnamefont {R.}~\bibnamefont
  {Rabindranath}}, \ and\ \bibinfo {author} {\bibfnamefont {J.}~\bibnamefont
  {Ehrlich}},\ }\href@noop {} {\bibfield  {journal} {\bibinfo  {journal} {J.
  Intell. Mater. Syst. Struct.}\ }\textbf {\bibinfo {volume} {23}},\ \bibinfo
  {pages} {989} (\bibinfo {year} {2012})}\BibitemShut {NoStop}%
\bibitem [{\citenamefont {S\"oderberg}\ \emph {et~al.}(2005)\citenamefont
  {S\"oderberg}, \citenamefont {Ge}, \citenamefont {Sozinov}, \citenamefont
  {Hannula},\ and\ \citenamefont {Lindroos}}]{Soederberg2005_SmartMaterStruct}%
  \BibitemOpen
  \bibfield  {author} {\bibinfo {author} {\bibfnamefont {O.}~\bibnamefont
  {S\"oderberg}}, \bibinfo {author} {\bibfnamefont {Y.}~\bibnamefont {Ge}},
  \bibinfo {author} {\bibfnamefont {A.}~\bibnamefont {Sozinov}}, \bibinfo
  {author} {\bibfnamefont {S.-P.}\ \bibnamefont {Hannula}}, \ and\ \bibinfo
  {author} {\bibfnamefont {V.~K.}\ \bibnamefont {Lindroos}},\ }\href@noop {}
  {\bibfield  {journal} {\bibinfo  {journal} {Smart Mater. Struct.}\ }\textbf
  {\bibinfo {volume} {14}},\ \bibinfo {pages} {S223} (\bibinfo {year}
  {2005})}\BibitemShut {NoStop}%
\bibitem [{\citenamefont {Karaca}\ \emph {et~al.}(2006)\citenamefont {Karaca},
  \citenamefont {Karaman}, \citenamefont {Basaran}, \citenamefont
  {Chumlyakov},\ and\ \citenamefont {Maier}}]{Karaca2006_ActaMater}%
  \BibitemOpen
  \bibfield  {author} {\bibinfo {author} {\bibfnamefont {H.~E.}\ \bibnamefont
  {Karaca}}, \bibinfo {author} {\bibfnamefont {I.}~\bibnamefont {Karaman}},
  \bibinfo {author} {\bibfnamefont {B.}~\bibnamefont {Basaran}}, \bibinfo
  {author} {\bibfnamefont {Y.~I.}\ \bibnamefont {Chumlyakov}}, \ and\ \bibinfo
  {author} {\bibfnamefont {H.~J.}\ \bibnamefont {Maier}},\ }\href@noop {}
  {\bibfield  {journal} {\bibinfo  {journal} {Acta Mater.}\ }\textbf {\bibinfo
  {volume} {54}},\ \bibinfo {pages} {233} (\bibinfo {year} {2006})}\BibitemShut
  {NoStop}%
\bibitem [{\citenamefont {El~Feninat}\ \emph {et~al.}(2002)\citenamefont
  {El~Feninat}, \citenamefont {Laroche}, \citenamefont {Fiset},\ and\
  \citenamefont {Mantovani}}]{ElFeninat2002_AdvEngMater}%
  \BibitemOpen
  \bibfield  {author} {\bibinfo {author} {\bibfnamefont {F.}~\bibnamefont
  {El~Feninat}}, \bibinfo {author} {\bibfnamefont {G.}~\bibnamefont {Laroche}},
  \bibinfo {author} {\bibfnamefont {M.}~\bibnamefont {Fiset}}, \ and\ \bibinfo
  {author} {\bibfnamefont {D.}~\bibnamefont {Mantovani}},\ }\href@noop {}
  {\bibfield  {journal} {\bibinfo  {journal} {Adv. Eng. Mater.}\ }\textbf
  {\bibinfo {volume} {4}},\ \bibinfo {pages} {91} (\bibinfo {year}
  {2002})}\BibitemShut {NoStop}%
\bibitem [{\citenamefont {Liu}, \citenamefont {Qin},\ and\ \citenamefont
  {Mather}(2007)}]{Liu2007_JMaterChem}%
  \BibitemOpen
  \bibfield  {author} {\bibinfo {author} {\bibfnamefont {C.}~\bibnamefont
  {Liu}}, \bibinfo {author} {\bibfnamefont {H.}~\bibnamefont {Qin}}, \ and\
  \bibinfo {author} {\bibfnamefont {P.~T.}\ \bibnamefont {Mather}},\
  }\href@noop {} {\bibfield  {journal} {\bibinfo  {journal} {J. Mater. Chem.}\
  }\textbf {\bibinfo {volume} {17}},\ \bibinfo {pages} {1543} (\bibinfo {year}
  {2007})}\BibitemShut {NoStop}%
\bibitem [{\citenamefont {Sokolowski}\ \emph {et~al.}(2007)\citenamefont
  {Sokolowski}, \citenamefont {Metcalfe}, \citenamefont {Hayashi},
  \citenamefont {Yahia},\ and\ \citenamefont
  {Raymond}}]{Sokolowski2007_BiomedMater}%
  \BibitemOpen
  \bibfield  {author} {\bibinfo {author} {\bibfnamefont {W.}~\bibnamefont
  {Sokolowski}}, \bibinfo {author} {\bibfnamefont {A.}~\bibnamefont
  {Metcalfe}}, \bibinfo {author} {\bibfnamefont {S.}~\bibnamefont {Hayashi}},
  \bibinfo {author} {\bibfnamefont {L.}~\bibnamefont {Yahia}}, \ and\ \bibinfo
  {author} {\bibfnamefont {J.}~\bibnamefont {Raymond}},\ }\href@noop {}
  {\bibfield  {journal} {\bibinfo  {journal} {Biomed. Mater.}\ }\textbf
  {\bibinfo {volume} {2}},\ \bibinfo {pages} {S23} (\bibinfo {year}
  {2007})}\BibitemShut {NoStop}%
\bibitem [{\citenamefont {Leng}\ \emph {et~al.}(2009)\citenamefont {Leng},
  \citenamefont {Lu}, \citenamefont {Liu}, \citenamefont {Huang},\ and\
  \citenamefont {Du}}]{Leng2009_MRSBull}%
  \BibitemOpen
  \bibfield  {author} {\bibinfo {author} {\bibfnamefont {J.}~\bibnamefont
  {Leng}}, \bibinfo {author} {\bibfnamefont {H.}~\bibnamefont {Lu}}, \bibinfo
  {author} {\bibfnamefont {Y.}~\bibnamefont {Liu}}, \bibinfo {author}
  {\bibfnamefont {W.~M.}\ \bibnamefont {Huang}}, \ and\ \bibinfo {author}
  {\bibfnamefont {S.}~\bibnamefont {Du}},\ }\href@noop {} {\bibfield  {journal}
  {\bibinfo  {journal} {MRS Bull.}\ }\textbf {\bibinfo {volume} {34}},\
  \bibinfo {pages} {848} (\bibinfo {year} {2009})}\BibitemShut {NoStop}%
\bibitem [{\citenamefont {Behl}, \citenamefont {Razzaq},\ and\ \citenamefont
  {Lendlein}(2010)}]{Behl2010_AdvMater}%
  \BibitemOpen
  \bibfield  {author} {\bibinfo {author} {\bibfnamefont {M.}~\bibnamefont
  {Behl}}, \bibinfo {author} {\bibfnamefont {M.~Y.}\ \bibnamefont {Razzaq}}, \
  and\ \bibinfo {author} {\bibfnamefont {A.}~\bibnamefont {Lendlein}},\
  }\href@noop {} {\bibfield  {journal} {\bibinfo  {journal} {Adv. Mater.}\
  }\textbf {\bibinfo {volume} {22}},\ \bibinfo {pages} {3388} (\bibinfo {year}
  {2010})}\BibitemShut {NoStop}%
\bibitem [{\citenamefont {Li}\ \emph {et~al.}(2013)\citenamefont {Li},
  \citenamefont {Huang}, \citenamefont {Zhang}, \citenamefont {Li},
  \citenamefont {Chen}, \citenamefont {Lu}, \citenamefont {Lu},\ and\
  \citenamefont {Xu}}]{Li2013_AdvFunctMater}%
  \BibitemOpen
  \bibfield  {author} {\bibinfo {author} {\bibfnamefont {Y.}~\bibnamefont
  {Li}}, \bibinfo {author} {\bibfnamefont {G.}~\bibnamefont {Huang}}, \bibinfo
  {author} {\bibfnamefont {X.}~\bibnamefont {Zhang}}, \bibinfo {author}
  {\bibfnamefont {B.}~\bibnamefont {Li}}, \bibinfo {author} {\bibfnamefont
  {Y.}~\bibnamefont {Chen}}, \bibinfo {author} {\bibfnamefont {T.}~\bibnamefont
  {Lu}}, \bibinfo {author} {\bibfnamefont {T.~J.}\ \bibnamefont {Lu}}, \ and\
  \bibinfo {author} {\bibfnamefont {F.}~\bibnamefont {Xu}},\ }\href@noop {}
  {\bibfield  {journal} {\bibinfo  {journal} {Adv. Funct. Mater.}\ }\textbf
  {\bibinfo {volume} {23}},\ \bibinfo {pages} {660} (\bibinfo {year}
  {2013})}\BibitemShut {NoStop}%
\bibitem [{\citenamefont {Mohr}\ \emph {et~al.}(2006)\citenamefont {Mohr},
  \citenamefont {Kratz}, \citenamefont {Weigel}, \citenamefont {Lucka-Gabor},
  \citenamefont {Moneke},\ and\ \citenamefont
  {Lendlein}}]{Mohr2006_ProcNatlAcadSciUSA}%
  \BibitemOpen
  \bibfield  {author} {\bibinfo {author} {\bibfnamefont {R.}~\bibnamefont
  {Mohr}}, \bibinfo {author} {\bibfnamefont {K.}~\bibnamefont {Kratz}},
  \bibinfo {author} {\bibfnamefont {T.}~\bibnamefont {Weigel}}, \bibinfo
  {author} {\bibfnamefont {M.}~\bibnamefont {Lucka-Gabor}}, \bibinfo {author}
  {\bibfnamefont {M.}~\bibnamefont {Moneke}}, \ and\ \bibinfo {author}
  {\bibfnamefont {A.}~\bibnamefont {Lendlein}},\ }\href@noop {} {\bibfield
  {journal} {\bibinfo  {journal} {Proc. Natl. Acad. Sci. USA}\ }\textbf
  {\bibinfo {volume} {103}},\ \bibinfo {pages} {3540} (\bibinfo {year}
  {2006})}\BibitemShut {NoStop}%
\bibitem [{\citenamefont {Collin}\ \emph {et~al.}(2003)\citenamefont {Collin},
  \citenamefont {Auernhammer}, \citenamefont {Gavat}, \citenamefont
  {Martinoty},\ and\ \citenamefont {Brand}}]{Collin2003_MacromolRapidCommun}%
  \BibitemOpen
  \bibfield  {author} {\bibinfo {author} {\bibfnamefont {D.}~\bibnamefont
  {Collin}}, \bibinfo {author} {\bibfnamefont {G.~K.}\ \bibnamefont
  {Auernhammer}}, \bibinfo {author} {\bibfnamefont {O.}~\bibnamefont {Gavat}},
  \bibinfo {author} {\bibfnamefont {P.}~\bibnamefont {Martinoty}}, \ and\
  \bibinfo {author} {\bibfnamefont {H.~R.}\ \bibnamefont {Brand}},\ }\href@noop
  {} {\bibfield  {journal} {\bibinfo  {journal} {Macromol. Rapid Commun.}\
  }\textbf {\bibinfo {volume} {24}},\ \bibinfo {pages} {737} (\bibinfo {year}
  {2003})}\BibitemShut {NoStop}%
\bibitem [{\citenamefont {Bohlius}, \citenamefont {Brand},\ and\ \citenamefont
  {Pleiner}(2004)}]{Bohlius2004_PhysRevE}%
  \BibitemOpen
  \bibfield  {author} {\bibinfo {author} {\bibfnamefont {S.}~\bibnamefont
  {Bohlius}}, \bibinfo {author} {\bibfnamefont {H.~R.}\ \bibnamefont {Brand}},
  \ and\ \bibinfo {author} {\bibfnamefont {H.}~\bibnamefont {Pleiner}},\
  }\href@noop {} {\bibfield  {journal} {\bibinfo  {journal} {Phys. Rev. E}\
  }\textbf {\bibinfo {volume} {70}},\ \bibinfo {pages} {061411} (\bibinfo
  {year} {2004})}\BibitemShut {NoStop}%
\bibitem [{\citenamefont {Zubarev}\ and\ \citenamefont
  {Iskakova}(2000)}]{Zubarev2000_PhysRevE}%
  \BibitemOpen
  \bibfield  {author} {\bibinfo {author} {\bibfnamefont {A.~Y.}\ \bibnamefont
  {Zubarev}}\ and\ \bibinfo {author} {\bibfnamefont {L.~Y.}\ \bibnamefont
  {Iskakova}},\ }\href@noop {} {\bibfield  {journal} {\bibinfo  {journal}
  {Phys. Rev. E}\ }\textbf {\bibinfo {volume} {61}},\ \bibinfo {pages} {5415}
  (\bibinfo {year} {2000})}\BibitemShut {NoStop}%
\bibitem [{\citenamefont {Hynninen}\ and\ \citenamefont
  {Dijkstra}(2005)}]{Hynninen2005_PhysRevLett}%
  \BibitemOpen
  \bibfield  {author} {\bibinfo {author} {\bibfnamefont {A.-P.}\ \bibnamefont
  {Hynninen}}\ and\ \bibinfo {author} {\bibfnamefont {M.}~\bibnamefont
  {Dijkstra}},\ }\href@noop {} {\bibfield  {journal} {\bibinfo  {journal}
  {Phys. Rev. Lett.}\ }\textbf {\bibinfo {volume} {94}},\ \bibinfo {pages}
  {138303} (\bibinfo {year} {2005})}\BibitemShut {NoStop}%
\bibitem [{\citenamefont {Auernhammer}, \citenamefont {Collin},\ and\
  \citenamefont {Martinoty}(2006)}]{Auernhammer2006_JChemPhys}%
  \BibitemOpen
  \bibfield  {author} {\bibinfo {author} {\bibfnamefont {G.~K.}\ \bibnamefont
  {Auernhammer}}, \bibinfo {author} {\bibfnamefont {D.}~\bibnamefont {Collin}},
  \ and\ \bibinfo {author} {\bibfnamefont {P.}~\bibnamefont {Martinoty}},\
  }\href@noop {} {\bibfield  {journal} {\bibinfo  {journal} {J. Chem. Phys}\
  }\textbf {\bibinfo {volume} {124}},\ \bibinfo {eid} {204907} (\bibinfo {year}
  {2006})}\BibitemShut {NoStop}%
\bibitem [{\citenamefont {Smallenburg}\ \emph {et~al.}(2012)\citenamefont
  {Smallenburg}, \citenamefont {Vutukuri}, \citenamefont {Imhof}, \citenamefont
  {van Blaaderen},\ and\ \citenamefont
  {Dijkstra}}]{Smallenburg2012_JPhysCondensMatter}%
  \BibitemOpen
  \bibfield  {author} {\bibinfo {author} {\bibfnamefont {F.}~\bibnamefont
  {Smallenburg}}, \bibinfo {author} {\bibfnamefont {H.~R.}\ \bibnamefont
  {Vutukuri}}, \bibinfo {author} {\bibfnamefont {A.}~\bibnamefont {Imhof}},
  \bibinfo {author} {\bibfnamefont {A.}~\bibnamefont {van Blaaderen}}, \ and\
  \bibinfo {author} {\bibfnamefont {M.}~\bibnamefont {Dijkstra}},\ }\href@noop
  {} {\bibfield  {journal} {\bibinfo  {journal} {J. Phys.: Condens. Matter}\
  }\textbf {\bibinfo {volume} {24}},\ \bibinfo {pages} {464113} (\bibinfo
  {year} {2012})}\BibitemShut {NoStop}%
\bibitem [{\citenamefont {Frickel}, \citenamefont {Messing},\ and\
  \citenamefont {Schmidt}(2011)}]{Frickel2011_JMaterChem}%
  \BibitemOpen
  \bibfield  {author} {\bibinfo {author} {\bibfnamefont {N.}~\bibnamefont
  {Frickel}}, \bibinfo {author} {\bibfnamefont {R.}~\bibnamefont {Messing}}, \
  and\ \bibinfo {author} {\bibfnamefont {A.~M.}\ \bibnamefont {Schmidt}},\
  }\href@noop {} {\bibfield  {journal} {\bibinfo  {journal} {J. Mater. Chem.}\
  }\textbf {\bibinfo {volume} {21}},\ \bibinfo {pages} {8466} (\bibinfo {year}
  {2011})}\BibitemShut {NoStop}%
\bibitem [{\citenamefont {N{\'e}el}(1949)}]{Neel1949_AnnGeopHys}%
  \BibitemOpen
  \bibfield  {author} {\bibinfo {author} {\bibfnamefont {L.}~\bibnamefont
  {N{\'e}el}},\ }\href@noop {} {\bibfield  {journal} {\bibinfo  {journal} {Ann.
  G{\'e}ophys}\ }\textbf {\bibinfo {volume} {5}},\ \bibinfo {pages} {99}
  (\bibinfo {year} {1949})}\BibitemShut {NoStop}%
\bibitem [{\citenamefont {Gundermann}\ and\ \citenamefont
  {Odenbach}(2014)}]{Gundermann2014_SmartMaterStruct}%
  \BibitemOpen
  \bibfield  {author} {\bibinfo {author} {\bibfnamefont {T.}~\bibnamefont
  {Gundermann}}\ and\ \bibinfo {author} {\bibfnamefont {S.}~\bibnamefont
  {Odenbach}},\ }\href@noop {} {\bibfield  {journal} {\bibinfo  {journal}
  {Smart Mater. Struct.}\ }\textbf {\bibinfo {volume} {23}},\ \bibinfo {pages}
  {105013} (\bibinfo {year} {2014})}\BibitemShut {NoStop}%
\bibitem [{\citenamefont {Liu}\ \emph {et~al.}(2012)\citenamefont {Liu},
  \citenamefont {Xu}, \citenamefont {Che}, \citenamefont {Chen}, \citenamefont
  {Liu},\ and\ \citenamefont {Xia}}]{Liu2012_JMaterChem}%
  \BibitemOpen
  \bibfield  {author} {\bibinfo {author} {\bibfnamefont {J.}~\bibnamefont
  {Liu}}, \bibinfo {author} {\bibfnamefont {J.}~\bibnamefont {Xu}}, \bibinfo
  {author} {\bibfnamefont {R.}~\bibnamefont {Che}}, \bibinfo {author}
  {\bibfnamefont {H.}~\bibnamefont {Chen}}, \bibinfo {author} {\bibfnamefont
  {Z.}~\bibnamefont {Liu}}, \ and\ \bibinfo {author} {\bibfnamefont
  {F.}~\bibnamefont {Xia}},\ }\href@noop {} {\bibfield  {journal} {\bibinfo
  {journal} {J. Mater. Chem.}\ }\textbf {\bibinfo {volume} {22}},\ \bibinfo
  {pages} {9277} (\bibinfo {year} {2012})}\BibitemShut {NoStop}%
\bibitem [{\citenamefont {Okada}\ \emph {et~al.}(2013)\citenamefont {Okada},
  \citenamefont {Nagao}, \citenamefont {Ueno}, \citenamefont {Ishii},\ and\
  \citenamefont {Konno}}]{Okada2013_Langmuir}%
  \BibitemOpen
  \bibfield  {author} {\bibinfo {author} {\bibfnamefont {A.}~\bibnamefont
  {Okada}}, \bibinfo {author} {\bibfnamefont {D.}~\bibnamefont {Nagao}},
  \bibinfo {author} {\bibfnamefont {T.}~\bibnamefont {Ueno}}, \bibinfo {author}
  {\bibfnamefont {H.}~\bibnamefont {Ishii}}, \ and\ \bibinfo {author}
  {\bibfnamefont {M.}~\bibnamefont {Konno}},\ }\href@noop {} {\bibfield
  {journal} {\bibinfo  {journal} {Langmuir}\ }\textbf {\bibinfo {volume}
  {29}},\ \bibinfo {pages} {9004} (\bibinfo {year} {2013})}\BibitemShut
  {NoStop}%
\bibitem [{\citenamefont {Hartmann}\ and\ \citenamefont
  {Neff}(2003)}]{Hartmann2003_IntJSolidsStruct}%
  \BibitemOpen
  \bibfield  {author} {\bibinfo {author} {\bibfnamefont {S.}~\bibnamefont
  {Hartmann}}\ and\ \bibinfo {author} {\bibfnamefont {P.}~\bibnamefont
  {Neff}},\ }\href@noop {} {\bibfield  {journal} {\bibinfo  {journal} {Int. J.
  Solids Struct.}\ }\textbf {\bibinfo {volume} {40}},\ \bibinfo {pages} {2767}
  (\bibinfo {year} {2003})}\BibitemShut {NoStop}%
\bibitem [{sup()}]{supplemental}%
  \BibitemOpen
  \href@noop {} {}\bibinfo {note} {See Supplemental Material at \textit{[URL
  will be inserted by AIP]} for further details on our numerical
  implementation, on the effect of the rigid inclusions, on our measurements of
  orientational order during a strain cycle, and for an energetic consideration
  of the flipping process on a cuboid lattice.}\BibitemShut {Stop}%
\bibitem [{\citenamefont {Geuzaine}\ and\ \citenamefont
  {Remacle}(2009)}]{Geuzaine2009_IntJNumerMethEng}%
  \BibitemOpen
  \bibfield  {author} {\bibinfo {author} {\bibfnamefont {C.}~\bibnamefont
  {Geuzaine}}\ and\ \bibinfo {author} {\bibfnamefont {J.-F.}\ \bibnamefont
  {Remacle}},\ }\href@noop {} {\bibfield  {journal} {\bibinfo  {journal} {Int.
  J. Numer. Meth. Eng.}\ }\textbf {\bibinfo {volume} {79}},\ \bibinfo {pages}
  {1309} (\bibinfo {year} {2009})}\BibitemShut {NoStop}%
\bibitem [{\citenamefont {Annunziata}, \citenamefont {Menzel},\ and\
  \citenamefont {L\"owen}(2013)}]{Annunziata2013_JChemPhys}%
  \BibitemOpen
  \bibfield  {author} {\bibinfo {author} {\bibfnamefont {M.~A.}\ \bibnamefont
  {Annunziata}}, \bibinfo {author} {\bibfnamefont {A.~M.}\ \bibnamefont
  {Menzel}}, \ and\ \bibinfo {author} {\bibfnamefont {H.}~\bibnamefont
  {L\"owen}},\ }\href@noop {} {\bibfield  {journal} {\bibinfo  {journal} {J.
  Chem. Phys.}\ }\textbf {\bibinfo {volume} {138}},\ \bibinfo {eid} {204906}
  (\bibinfo {year} {2013})}\BibitemShut {NoStop}%
\bibitem [{\citenamefont {Biller}, \citenamefont {Stolbov},\ and\ \citenamefont
  {Raikher}(2014)}]{Biller2014_JApplPhys}%
  \BibitemOpen
  \bibfield  {author} {\bibinfo {author} {\bibfnamefont {A.~M.}\ \bibnamefont
  {Biller}}, \bibinfo {author} {\bibfnamefont {O.~V.}\ \bibnamefont {Stolbov}},
  \ and\ \bibinfo {author} {\bibfnamefont {Y.~L.}\ \bibnamefont {Raikher}},\
  }\href@noop {} {\bibfield  {journal} {\bibinfo  {journal} {J. Appl. Phys.}\
  }\textbf {\bibinfo {volume} {116}},\ \bibinfo {pages} {114904} (\bibinfo
  {year} {2014})}\BibitemShut {NoStop}%
\bibitem [{\citenamefont {Menzel}(2014)}]{Menzel2014_JChemPhys}%
  \BibitemOpen
  \bibfield  {author} {\bibinfo {author} {\bibfnamefont {A.~M.}\ \bibnamefont
  {Menzel}},\ }\href@noop {} {\bibfield  {journal} {\bibinfo  {journal} {J.
  Chem. Phys.}\ }\textbf {\bibinfo {volume} {141}},\ \bibinfo {eid} {194907}
  (\bibinfo {year} {2014})}\BibitemShut {NoStop}%
\bibitem [{\citenamefont {Chernenko}\ \emph {et~al.}(2003)\citenamefont
  {Chernenko}, \citenamefont {L'vov}, \citenamefont {Pons},\ and\ \citenamefont
  {Cesari}}]{Chernenko2003_JApplPhys}%
  \BibitemOpen
  \bibfield  {author} {\bibinfo {author} {\bibfnamefont {V.~A.}\ \bibnamefont
  {Chernenko}}, \bibinfo {author} {\bibfnamefont {V.}~\bibnamefont {L'vov}},
  \bibinfo {author} {\bibfnamefont {J.}~\bibnamefont {Pons}}, \ and\ \bibinfo
  {author} {\bibfnamefont {E.}~\bibnamefont {Cesari}},\ }\href@noop {}
  {\bibfield  {journal} {\bibinfo  {journal} {J. Appl. Phys.}\ }\textbf
  {\bibinfo {volume} {93}},\ \bibinfo {pages} {2394} (\bibinfo {year}
  {2003})}\BibitemShut {NoStop}%
\bibitem [{\citenamefont {Arghavani}, \citenamefont {Auricchio},\ and\
  \citenamefont {Naghdabadi}(2011)}]{Arghavani2011_IntJPlasticity}%
  \BibitemOpen
  \bibfield  {author} {\bibinfo {author} {\bibfnamefont {J.}~\bibnamefont
  {Arghavani}}, \bibinfo {author} {\bibfnamefont {F.}~\bibnamefont
  {Auricchio}}, \ and\ \bibinfo {author} {\bibfnamefont {R.}~\bibnamefont
  {Naghdabadi}},\ }\href@noop {} {\bibfield  {journal} {\bibinfo  {journal}
  {Int. J. Plasticity}\ }\textbf {\bibinfo {volume} {27}},\ \bibinfo {pages}
  {940} (\bibinfo {year} {2011})}\BibitemShut {NoStop}%
\bibitem [{\citenamefont {Borb\'ath}\ \emph {et~al.}(2012)\citenamefont
  {Borb\'ath}, \citenamefont {G\"unther}, \citenamefont {Borin}, \citenamefont
  {Gundermann},\ and\ \citenamefont {Odenbach}}]{Borbath2012_SmartMaterStruct}%
  \BibitemOpen
  \bibfield  {author} {\bibinfo {author} {\bibfnamefont {T.}~\bibnamefont
  {Borb\'ath}}, \bibinfo {author} {\bibfnamefont {S.}~\bibnamefont
  {G\"unther}}, \bibinfo {author} {\bibfnamefont {D.~Y.}\ \bibnamefont
  {Borin}}, \bibinfo {author} {\bibfnamefont {T.}~\bibnamefont {Gundermann}}, \
  and\ \bibinfo {author} {\bibfnamefont {S.}~\bibnamefont {Odenbach}},\
  }\href@noop {} {\bibfield  {journal} {\bibinfo  {journal} {Smart Mater.
  Struct.}\ }\textbf {\bibinfo {volume} {21}},\ \bibinfo {pages} {105018}
  (\bibinfo {year} {2012})}\BibitemShut {NoStop}%
\bibitem [{\citenamefont {Kramarenko}\ \emph {et~al.}(2015)\citenamefont
  {Kramarenko}, \citenamefont {Chertovich}, \citenamefont {Stepanov},
  \citenamefont {Semisalova}, \citenamefont {Makarova}, \citenamefont {Perov},\
  and\ \citenamefont {Khokhlov}}]{Kramarenko2015_SmartMaterStruct}%
  \BibitemOpen
  \bibfield  {author} {\bibinfo {author} {\bibfnamefont {E.~Y.}\ \bibnamefont
  {Kramarenko}}, \bibinfo {author} {\bibfnamefont {A.~V.}\ \bibnamefont
  {Chertovich}}, \bibinfo {author} {\bibfnamefont {G.~V.}\ \bibnamefont
  {Stepanov}}, \bibinfo {author} {\bibfnamefont {A.~S.}\ \bibnamefont
  {Semisalova}}, \bibinfo {author} {\bibfnamefont {L.~A.}\ \bibnamefont
  {Makarova}}, \bibinfo {author} {\bibfnamefont {N.~S.}\ \bibnamefont {Perov}},
  \ and\ \bibinfo {author} {\bibfnamefont {A.~R.}\ \bibnamefont {Khokhlov}},\
  }\href@noop {} {\bibfield  {journal} {\bibinfo  {journal} {Smart Mater.
  Struct.}\ }\textbf {\bibinfo {volume} {24}},\ \bibinfo {pages} {035002}
  (\bibinfo {year} {2015})}\BibitemShut {NoStop}%
\bibitem [{\citenamefont {Hoang}, \citenamefont {Zhang},\ and\ \citenamefont
  {Du}(2009)}]{Hoang2009_SmartMaterStruct}%
  \BibitemOpen
  \bibfield  {author} {\bibinfo {author} {\bibfnamefont {N.}~\bibnamefont
  {Hoang}}, \bibinfo {author} {\bibfnamefont {N.}~\bibnamefont {Zhang}}, \ and\
  \bibinfo {author} {\bibfnamefont {N.}~\bibnamefont {Du}},\ }\href@noop {}
  {\bibfield  {journal} {\bibinfo  {journal} {Smart Mater. Struct.}\ }\textbf
  {\bibinfo {volume} {18}},\ \bibinfo {pages} {074009} (\bibinfo {year}
  {2009})}\BibitemShut {NoStop}%
\bibitem [{\citenamefont {Chertovich}\ \emph {et~al.}(2010)\citenamefont
  {Chertovich}, \citenamefont {Stepanov}, \citenamefont {Kramarenko},\ and\
  \citenamefont {Khokhlov}}]{Chertovich2010_MacromolMaterEng}%
  \BibitemOpen
  \bibfield  {author} {\bibinfo {author} {\bibfnamefont {A.~V.}\ \bibnamefont
  {Chertovich}}, \bibinfo {author} {\bibfnamefont {G.~V.}\ \bibnamefont
  {Stepanov}}, \bibinfo {author} {\bibfnamefont {E.~Y.}\ \bibnamefont
  {Kramarenko}}, \ and\ \bibinfo {author} {\bibfnamefont {A.~R.}\ \bibnamefont
  {Khokhlov}},\ }\href@noop {} {\bibfield  {journal} {\bibinfo  {journal}
  {Macromol. Mater. Eng.}\ }\textbf {\bibinfo {volume} {295}},\ \bibinfo
  {pages} {336} (\bibinfo {year} {2010})}\BibitemShut {NoStop}%
\bibitem [{\citenamefont {Huang}\ \emph {et~al.}(2015)\citenamefont {Huang},
  \citenamefont {Pessot}, \citenamefont {Cremer}, \citenamefont {Weeber},
  \citenamefont {Holm}, \citenamefont {Nowak}, \citenamefont {Odenbach},
  \citenamefont {Menzel},\ and\ \citenamefont
  {Auernhammer}}]{Huang2015_SoftMatter}%
  \BibitemOpen
  \bibfield  {author} {\bibinfo {author} {\bibfnamefont {S.}~\bibnamefont
  {Huang}}, \bibinfo {author} {\bibfnamefont {G.}~\bibnamefont {Pessot}},
  \bibinfo {author} {\bibfnamefont {P.}~\bibnamefont {Cremer}}, \bibinfo
  {author} {\bibfnamefont {R.}~\bibnamefont {Weeber}}, \bibinfo {author}
  {\bibfnamefont {C.}~\bibnamefont {Holm}}, \bibinfo {author} {\bibfnamefont
  {J.}~\bibnamefont {Nowak}}, \bibinfo {author} {\bibfnamefont
  {S.}~\bibnamefont {Odenbach}}, \bibinfo {author} {\bibfnamefont {A.~M.}\
  \bibnamefont {Menzel}}, \ and\ \bibinfo {author} {\bibfnamefont {G.~K.}\
  \bibnamefont {Auernhammer}},\ }\href@noop {} {\bibfield  {journal} {\bibinfo
  {journal} {Soft Matter}\ } (\bibinfo {year} {2015})},\ \bibinfo {note} {{DOI:
  10.1039/C5SM01814E }}\BibitemShut {NoStop}%
\end{thebibliography}


\end{document}